\documentclass[aps,prl,twocolumn,superscriptaddress,longbibliography]{revtex4-2}
\pdfoutput=1
\usepackage{graphicx}
\usepackage{amssymb,amsmath}
\usepackage{bm,dsfont}
\usepackage{dcolumn}
\usepackage{float}
\usepackage[OT1]{fontenc} 
\usepackage{url}
\usepackage{mathrsfs}
\usepackage{slashed,comment}
\usepackage{color}
\usepackage{verbatim}
\usepackage{enumitem}
\usepackage{soul,physics}
\usepackage[driverfallback=dvipdfm]{hyperref}
\hypersetup{pdfpagemode=FullScreen,colorlinks=true,breaklinks,urlcolor=blue,linkcolor=blue,citecolor=blue}

\usepackage{subfigure}
\usepackage{amssymb}
\usepackage{bm}
\usepackage{graphicx}
\usepackage{amsmath,amssymb}
\usepackage{xcolor}
\usepackage{tikz,fp}
\usepackage{tikz-cd}
\usetikzlibrary{arrows}
\usetikzlibrary{intersections}
\usetikzlibrary{shapes.geometric}
\usetikzlibrary{decorations.pathmorphing, patterns,shapes,fixedpointarithmetic}
\usetikzlibrary{decorations.markings}

\begin{document}
	
	\title{Bridging Krylov Complexity and Universal Analog Quantum Simulator}
	
    \author{Shuo Zhang}
    \thanks{These authors contributed equally to this work.}
    \affiliation{State Key Laboratory of Surface Physics \& Department of Physics, Fudan University, Shanghai, 200438, China}

    \author{Yuzhi Tong}
    \thanks{These authors contributed equally to this work.}
    \affiliation{State Key Laboratory of Surface Physics \& Department of Physics, Fudan University, Shanghai, 200438, China}
    \affiliation{Institute for Advanced Study, Tsinghua University, Beijing, 100084, China}

    \author{Pengfei Zhang}
    \thanks{PengfeiZhang.physics@gmail.com}
    \affiliation{State Key Laboratory of Surface Physics \& Department of Physics, Fudan University, Shanghai, 200438, China}
    \affiliation{Hefei National Laboratory, Hefei 230088, China}

    \author{Zeyu Liu}
    \thanks{zeyuliuphysics@gmail.com}
    \affiliation{State Key Laboratory of Surface Physics \& Department of Physics, Fudan University, Shanghai, 200438, China}

	\date{\today}
	\begin{abstract}
    Quantum simulation of complex many-body systems beyond classical computational capabilities provides a promising route toward understanding novel quantum phases and their transitions. In particular, analog quantum simulators with global control fields have attracted considerable attention due to their potential to simulate arbitrary Hamiltonians and perform quantum computing tasks. However, a clear, quantitative measure for the complexity of implementing specific quantum operations in such systems is still lacking. In this Letter, we address this challenge by introducing generalized Krylov complexity, a concept originating from operator growth dynamics, as a direct diagnosis for this synthesis complexity. We construct the block Krylov basis generated by a set of Hamiltonians, which naturally organizes the operator space achievable through the simulator’s native interactions and their nested commutators. By analyzing representative systems including Rydberg atom arrays, we demonstrate that the generalized Krylov complexity of a target operation serves as a strong predictor of the minimum time required for its realization. Our results establish Krylov complexity as an intuitive and predictive tool for designing efficient control protocols in analog quantum simulators.
	\end{abstract}
	
	\maketitle

    \emph{ \color{blue} Introduction.--} Quantum complexity quantifies the minimal resources required to implement a specific quantum operation \cite{Roberts:2016hpo, Jefferson:2017sdb, Yang:2017nfn, Chapman:2017rqy, Khan:2018rzm, Yang:2018nda, Lucas:2018wsc, Balasubramanian:2019wgd, Balasubramanian:2021mxo}. This concept lies at the foundation of quantum information and computation, serving as a measure of algorithmic efficiency \cite{Nielsen:2012yss, Barenco:1995na, Deutsch:1995dw, Lloyd:1996aai, Bremner:2002xdg}. It also plays a significant role in broader physical contexts, notably in the study of black holes and holographic quantum systems \cite{Susskind:2014rva,Stanford:2014jda,Brown:2015bva,Brown:2015lvg,Caputa:2017urj,Caputa:2017yrh,Brown:2019rox,Belin:2021bga,Belin:2022xmt,Miyaji:2015woj,Miyaji:2016fse,Belin:2018bpg,Brown:2017jil,Haferkamp:2021uxo}. Among the various notions of quantum complexity, quantum circuit complexity has been extensively studied. It provides a concrete, operational measure for the complexity of a target unitary operation in terms of the minimal number of basic building blocks required to approximate it: Given a target unitary operator $\hat{\mathcal{U}}$, the goal is to approximate it using a sequence of elementary quantum gates $\{ \hat{U}_i \}$, such that $ \hat{\mathcal{U}} \approx  \hat{U}_1 \cdots \hat{U}_n.$ The quantum circuit complexity of $\hat{\mathcal{U}}$ is then defined as the smallest integer $n$ for which $ \| \hat{\mathcal{U}} - \hat{U}_1 \cdots \hat{U}_n \| < \varepsilon,$ given a prescribed tolerance $\varepsilon > 0$ and an appropriate operator norm.


    Recent developments have identified analog quantum simulators with global control fields as a powerful paradigm for implementing complex quantum dynamics \cite{Chiu:2025uis, Manetsch:2024lwl, Guo:2024caa, Semeghini:2021wls, Bohnet:2016qql, Scholl:2020hzx, Kornjaca:2024afu, Manovitz:2024hif, 2023Sci...381...82H, Andersen:2024aob, PhysRevLett.130.010201, Liang:2024bqn, Xu:2025nvo, mazurenko2017cold, 2022NatCo..13.6824W, 2016PhRvX...6b1030H, Shao:2024cop, 2017Sci...357..995G, Sompet:2021zow, Yue:2025zqf, Lu:2024guw, deLeseleuc:2018wyc, Gonzalez-Cuadra:2024xul, Mildenberger:2022jqr,Mark:2024gzn}, with the capacity for universal quantum computation \cite{Benjamin:1999dlz, 2001PhRvL..88a7904B, Lloyd:2018fsu, Cesa:2023hgs,Oszmaniec:2017cyf,Hu:2025omd}. The guiding principle is that if the dynamics generated by the native Hamiltonians $\hat{H}_{1}$ and $\hat{H}_{2}$ are experimentally accessible, then their linear combinations $\alpha \hat{H}_{1}+\beta \hat{H}_{2}$ and their commutator $i[\hat{H}_{1},\hat{H}_{2}]$ can also be effectively realized through specific pulse sequences. Universality is achieved when the \textit{dynamical Lie algebra} generated by these native interactions spans the entire operator space \cite{DAlessandro:2009btr, d2021introduction, Khaneja:2001kpd, Hu:2025omd, Ragone:2023qbn, Wiersema:2023txu, Allcock:2024mgx,AraizaBravo:2024lcg}. Nevertheless, achieving universality does not guarantee the efficiency of realizing arbitrary evolutions. For a given set of native Hamiltonians, understanding which classes of operations can be implemented efficiently, or with low quantum circuit complexity, is therefore of fundamental significance.

     \begin{figure}[t]
        \centering
        \includegraphics[width=1\linewidth]{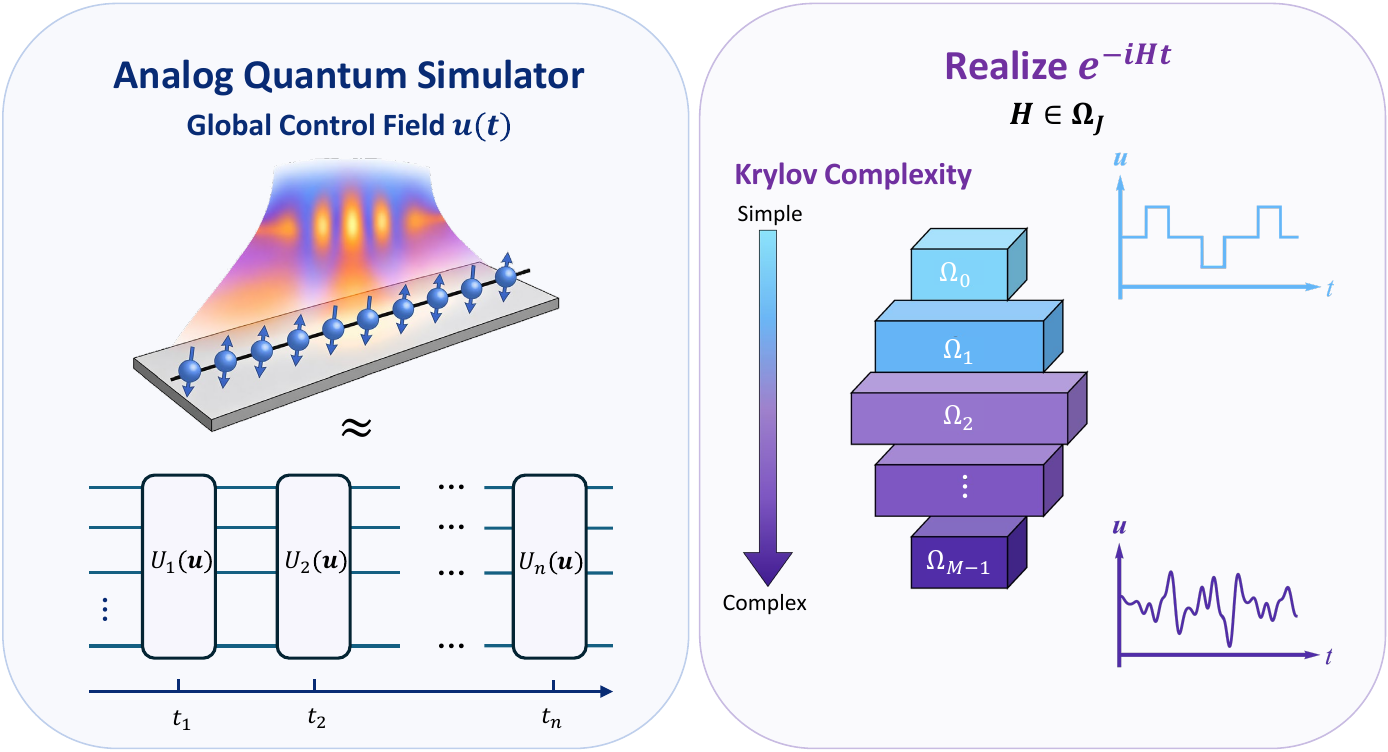}
        \caption{Schematic of a quantum simulator under global control fields and its associated block Krylov basis $\Omega_{m}$. For a target operation residing in the deeper Krylov subspace (larger $m$), the required pulse sequence $\bm{u}(t)$ becomes increasingly complicated. Consequently, the growth of Krylov complexity qualitatively mirrors that of quantum circuit complexity.}
        \label{fig:schemticas}
    \end{figure}
    
    In this Letter, we highlight a key observation: The structure of the \textit{dynamical Lie algebra} naturally suggests a link to Krylov complexity \cite{Parker:2018yvk, Avdoshkin:2022xuw, Balasubramanian:2022tpr, Liu:2022god, Barbon:2019wsy, Dymarsky:2019elm, Barbon:2019tuq, Magan:2020iac, Jian:2020qpp, Rabinovici:2020ryf, Chen:2019klo, Noh:2021vnc, 2021JHEP...12..188C, Patramanis:2021lkx, Bhattacharjee:2022vlt, Caputa:2021sib, Dymarsky:2021bjq, Avdoshkin:2019trj, Rabinovici:2023yex, Bhattacharya:2022gbz, Bhattacharjee:2022lzy, Bhattacharjee:2023uwx, Lv:2023jbv, Tang:2023ocr, Zhang:2023wtr,Craps:2024suj,Nandy:2024evd,Tan:2024kqb,Rabinovici:2025otw}, which quantifies the spread of an operator within the Krylov basis generated by repeated commutations with the system's native Hamiltonians. Crucially, the experimental generation of effective interactions through nested commutators in these platforms directly parallels the iterative construction of the Krylov basis. Motivated by this structural congruence, we introduce a generalized Krylov complexity in the context of universal quantum simulation by constructing a block Krylov basis generated by a set of native Hamiltonians. For a range of examples, including Rydberg atom arrays, we compute the Krylov complexity for the target Hamiltonian and numerically estimate the quantum circuit complexity by minimizing the evolution time via gradient-based optimization. The results establish the generalized Krylov complexity as a robust theoretical proxy for the quantum circuit complexity required to realize a target operation in analog quantum simulators. Our framework provides a concrete metric to benchmark and distinguish the performance of diverse analog quantum simulators.

    \emph{ \color{blue} Setup.--} We consider an analog quantum simulator with global control fields, as illustrated in Fig.~\ref{fig:schemticas}. Specifically, the system consists of $N$ qubits governed by the time-dependent Hamiltonian $\hat{H}(t)=\sum_{\alpha}u_{\alpha}(t)\hat{H}_{\alpha}$, where $\{\hat{H}_{\alpha}\}$ denotes the set of native Hamiltonians accessible to the simulator, and $\bm{u}(t)=\{u_{\alpha}(t)\}$ represents the corresponding set of time-dependent control pulses. The resulting time-evolution operator is given by
\begin{equation}\label{eq:Ugeneral}
\begin{aligned}
	\hat{\mathcal{U}}(\tau)=\mathcal{T}\left[ \exp\left(-i\int_{0}^{\tau}\dd t~\sum_{\alpha}u_{\alpha}(t)\hat{H}_{\alpha}\right) \right],
\end{aligned}
\end{equation}
    where $\mathcal{T}$ is the time-ordering operator.

    The time-dependence of the control fields enables the realization of evolutions generated by a large class of effective Hamiltonians. If the simulator can implement the dynamics generated by $\hat{H}_{1}$ and $\hat{H}_{2}$, it can also approximate the evolution under their linear combination $\alpha\hat{H}_{1}+\beta\hat{H}_{2}$ via the Trotter formula \cite{Suzuki:1976be, 1977CMaPh..57..193S}
\begin{equation}
\begin{aligned}
	e^{i(\alpha \hat{H}_{1}+\beta \hat{H}_{2})}=\lim_{n\to\infty}\left( e^{i\alpha \hat{H}_{1}/n}e^{i\beta \hat{H}_{2}/n} \right)^{n}.
\end{aligned}
\end{equation}
    Moreover, the effective evolution generated by the commutator of these operators can be synthesized using 
\begin{equation}\label{commute}
\begin{aligned}
	&e^{-[\hat{H}_{1},\hat{H}_{2}]}=\\
    &\lim_{n\to\infty}\left( e^{i\hat{H}_{1}/\sqrt{n}}e^{i\hat{H}_{2}/\sqrt{n}}e^{-i\hat{H}_{1}/\sqrt{n}}e^{-i\hat{H}_{2}/\sqrt{n}} \right)^{n},
\end{aligned}
\end{equation}
    which follows from the Baker-Campbell-Hausdorff formula \cite{Hall:2015xtd, rossmann2006lie}. Therefore, by appropriately designing the control pulse sequences, one can generate operations corresponding to arbitrary linear combinations and nested commutators of the native Hamiltonian set $\{ \hat{H}_{\alpha} \}$. The algebra generated through such repeated commutations is known as the \textit{dynamical Lie algebra} \cite{DAlessandro:2009btr, d2021introduction, Khaneja:2001kpd, Hu:2025omd, Ragone:2023qbn, Wiersema:2023txu, Allcock:2024mgx,AraizaBravo:2024lcg}. Universality is achieved when this algebra spans the entire operator space (except for the identity operator).

    \emph{ \color{blue} Generalized Krylov complexity.--} Motivated by discussions in the introduction, we turn to a generalized Krylov basis construction to quantify the complexity of exploring this algebraic space. In the traditional framework \cite{Parker:2018yvk}, an operator $\hat{\mathcal{O}}=\sum_{ij}\mathcal{O}_{ij}|i\rangle\langle j|$ is mapped to a state in the doubled Hilbert space, denoted by $|\hat{\mathcal{O}}\rangle=\sum_{ij}\mathcal{O}_{ij}|i\rangle\otimes |j\rangle$. The Liouvillian superoperator is defined by its action $\hat{\mathcal{L}}|\hat{X}\rangle=|[\hat{H},\hat{X}]\rangle$. The Krylov basis is then constructed by repeatedly applying the Liouvillian to an initial operator, and the Krylov complexity is defined to characterize the depth of this Krylov space exploration \cite{Parker:2018yvk, Avdoshkin:2022xuw, Balasubramanian:2022tpr, Liu:2022god, Barbon:2019wsy, Dymarsky:2019elm, Barbon:2019tuq, Magan:2020iac, Jian:2020qpp, Rabinovici:2020ryf, Chen:2019klo, Noh:2021vnc, 2021JHEP...12..188C, Patramanis:2021lkx, Bhattacharjee:2022vlt, Caputa:2021sib, Dymarsky:2021bjq, Avdoshkin:2019trj, Rabinovici:2023yex, Bhattacharya:2022gbz, Bhattacharjee:2022lzy, Bhattacharjee:2023uwx, Lv:2023jbv, Tang:2023ocr, Zhang:2023wtr,Craps:2024suj,Nandy:2024evd,Tan:2024kqb,Rabinovici:2025otw}. Its relation to circuit complexity in systems evolved under group generators has also been explored in \cite{Lv:2023jbv}.

    Our construction extends this setup in two aspects. First, the starting point is generalized from a single operator to the subspace $\Omega_{0}=\text{span}\{|\hat{H}_{\alpha}\rangle\}$ spanned by all native Hamiltonians available on the simulator. Second, instead of a single Liouvillian, we employ a collection of superoperators $\{\hat{\mathcal{L}}_{\alpha}\}$ defined by their actions $\hat{\mathcal{L}}_{\alpha}|\hat{X}\rangle=|[\hat{H}_{\alpha},\hat{X}]\rangle$. By iteratively applying these superoperators to the initial subspace, we generate the sequence
\begin{equation}
    \{ \Omega_{0}, \hat{\bm{\mathcal{L}}}\Omega_{0}, \ldots, \hat{\bm{\mathcal{L}}}^{n}\Omega_{0}, \ldots \},
\end{equation}
    where $\hat{\bm{\mathcal{L}}}$ denotes the collective action of all $\hat{\mathcal{L}}_{\alpha}$. This sequence, however, is generally neither orthogonal nor normalized. Therefore, we perform the Gram-Schmidt orthogonalization procedure using the Hilbert-Schmidt inner product $\langle\hat{\mathcal{A}}|\hat{\mathcal{B}}\rangle=\text{tr}[\hat{\mathcal{A}}^{\dagger}\hat{\mathcal{B}}]/\mathcal{D}$, where $\mathcal{D}$ is the Hilbert space dimension. The result is the block Krylov basis $\{\Omega_{0},\Omega_{1},\cdots,\Omega_{\mathcal{M}-1}\}$, where $\mathcal{M}$ is the total number of blocks. If a block $\Omega_{J}$ contains $n_{J}$ linearly independent operators, it is represented by the orthonormal set
\begin{equation}
    \Omega_{J}=\{{|\hat{\Omega}_{J,0}\rangle,\cdots, |\hat{\Omega}_{J,n_{J}-1}\rangle}\}.
\end{equation}

    To formalize the resource cost, we define the layer circuit complexity $C_J$ as the characteristic complexity of operators within the $J$-th layer $\Omega_J$. For any individual operator $\hat{A}$, its quantum circuit complexity $C(\hat{A})$ is defined as the minimum number of fundamental unitary evolutions generated by the base set $\Omega_0$ (with $C_0 \equiv 1$) required to realize the evolution $e^{i\hat{A}}$. Accordingly, we take $C_J = \max_{|\hat{A}\rangle \in \Omega_J} C(\hat{A})$. Since an operator in $\Omega_{J+1}$ is generated via the commutator $i[\hat{H}_1, \hat{H}_2]$ with $\hat{H}_1 \in \Omega_J$ and $\hat{H}_2 \in \Omega_0$, this necessitates a recursive overhead $C_{J+1} = 2C_J + 2$ as seen from Eq.~\eqref{commute}. Solving this recurrence relation yields $C_J = 3 \cdot 2^{J} - 2$, demonstrating an asymptotic exponential scaling $C_J \sim 2^J$. Building on this, we propose the generalized Krylov complexity $\mathcal{K}$ for a target Hamiltonian $\hat{H}_{\text{target}}$ as
\begin{equation}
\begin{aligned}
	P_{J}&=\sum_{m}\frac{|\langle \hat{\Omega}_{J,m}|\hat{H}_{\text{target}}\rangle|^{2}}{\langle \hat{H}_{\text{target}}|\hat{H}_{\text{target}}\rangle},\ \ \ \ \ \ \mathcal{K}&=\sum_{J}P_{J}~2^{J}.
\end{aligned}
\end{equation}
    Here, $P_J$ quantifies the distribution of the target Hamiltonian in the block Krylov basis. Compared to the traditional definition, we define the Krylov complexity using an exponential weighting $2^J$ instead of a linear function. We expect this generalized Krylov complexity to exhibit behavior consistent with the growth of the underlying quantum circuit complexity.

    \emph{ \color{blue} Models for quantum simulators.--} To concretely benchmark our framework, we consider a one-dimensional chain of $L$ qubits with open boundary conditions, governed by the native Hamiltonian set 
\begin{equation}
	\Omega_{0}=\text{span}\{ |\hat{H}_{X}\rangle,|\hat{H}_{Z}\rangle, |\hat{H}_{\text{break}}\rangle, |\hat{H}_{\text{int}}\rangle \}.
\end{equation}
    The single-body terms are defined as
\begin{equation}
    \hat{H}_{X}=\sum_{i=1}^{L}\hat{X}_{i},\ \ \ \ \ \hat{H}_{Z}=\sum_{i=1}^{L}\hat{Z}_{i},\ \ \ \ \ \hat{H}_{\text{break}}=\hat{X}_{1},
\end{equation}
	where $\hat{X}_i$ and $\hat{Z}_i$ denote the Pauli operators acting on site $i$. $\hat{H}_{\text{break}}$ breaks spatial reflection symmetry, which is necessary for achieving universal simulation capability \cite{Hu:2025omd}. To demonstrate the versatility of our framework, we specify $\hat{H}_{\text{int}}$ for two paradigmatic spin systems. The first case is the Ising coupling:
\begin{equation}
    \hat{H}_{\text{Ising}}=\sum_{i=1}^{L-1}\hat{Z}_{i}\hat{Z}_{i+1},
\end{equation}
    which provides an effective description for Rydberg atom arrays when long-range couplings can be safely neglected \cite{Jaksch:2000cus,RevModPhys.82.2313,2010NatPh...6..382W,Labuhn_2016,Bernien:2017ubn,2020NatPh..16..132B,Ebadi:2020ldi,Scholl:2020hzx,Wu:2020axb,Semeghini:2021wls,fang2024probingcriticalphenomenaopen,Cheng:2024nep,Bluvstein_2021,Morgado:2020jfo,Bluvstein:2021jsq,Ma:2023ltx,Scholl:2023cjt,Singh:2022qfv,Bluvstein:2023zmt,zhang2025observationnearcriticalkibblezurekscaling,Geim:2026bmr}. The second model we investigate is the isotropic Heisenberg model, defined as:
\begin{equation}
\begin{aligned}
    \hat{H}_{\text{Heisenberg}}=\sum_{i=1}^{L-1}\left(\hat{X}_{i}\hat{X}_{i+1}+\hat{Y}_{i}\hat{Y}_{i+1}+\hat{Z}_{i}\hat{Z}_{i+1}\right),
\end{aligned}
\end{equation}
    which emerges as the effective low-energy description of the Fermi Hubbard model, a paradigm that is naturally realized using ultracold atoms in optical lattices \cite{Trotzky:2008aor,Greif:2013byo,Parsons:2016ovf,Cheuk:2016dfi,mazurenko2017cold,shao2024antiferromagnetic}.

\begin{figure}[t]
        \centering
        \includegraphics[width=1\linewidth]{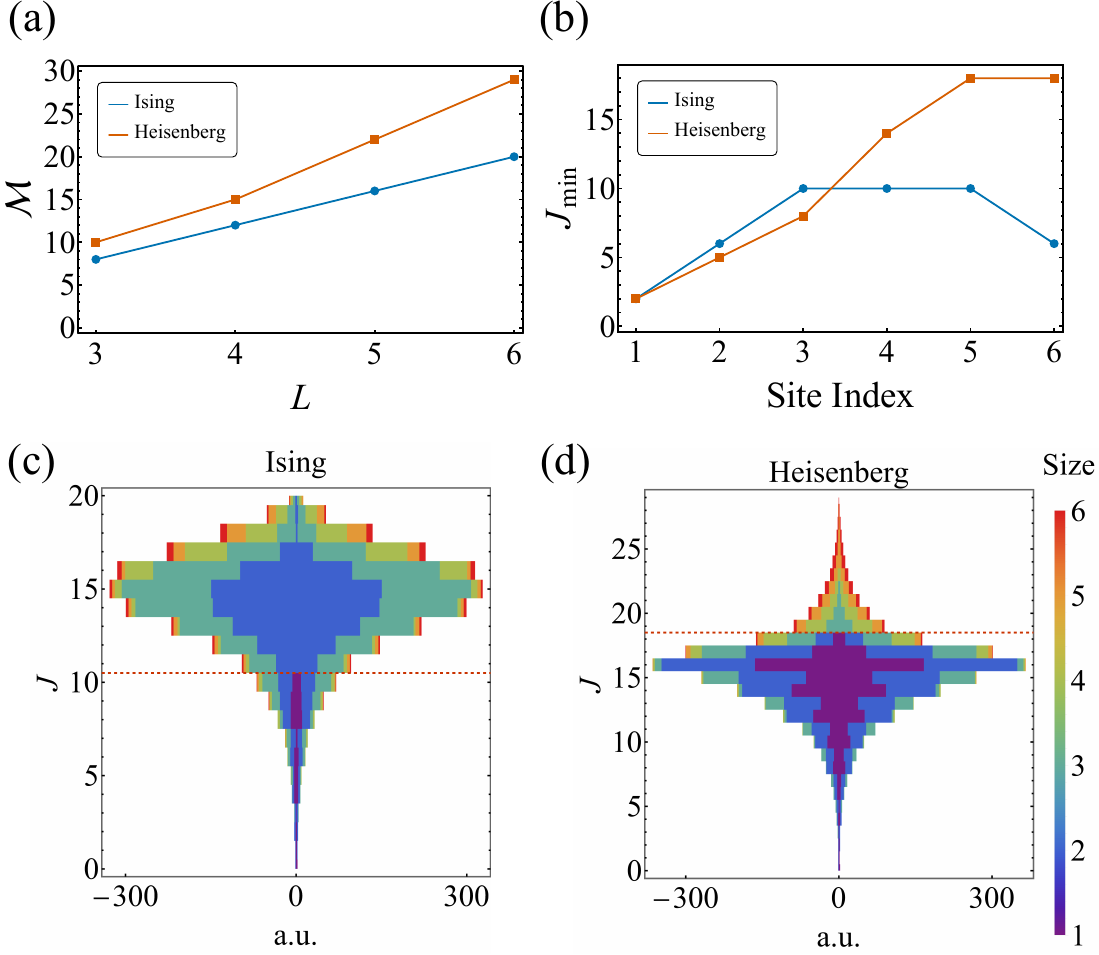}
        \caption{Structural properties of the block Krylov basis and operator spreading. (a) Maximum Krylov depth $\mathcal{M}$ (total number of blocks) as a function of system size $L$ for the Ising and Heisenberg models. (b) The minimum Krylov depth $J$ required to access all single-qubit Pauli operator ($\hat{X}_i$, $\hat{Y}_i$, and $\hat{Z}_i$). (c)-(d) Minimum operator size $S_{\text{min}}$ versus Krylov depth $J$ and intra-layer index for the (c) Ising and (d) Heisenberg models. The red dashed lines indicate the boundary below which all single-body operators ($S_{\text{min}}=1$) are contained.}
        \label{fig:2}
    \end{figure}

    For system sizes $L\in\{3,4,5,6\}$, we numerically verify that the \textit{dynamical Lie algebra} generated by the native Hamiltonian set spans the entire space of traceless operators, thereby indicating that both models facilitate universal quantum computation. The identity operator is inherently excluded, as the commutator of any two operators strictly yields a traceless generator. This numerical result is consistent with established universality proofs for the Ising model \cite{Hu:2025omd}.
    
    Fig.~\ref{fig:2}(a) illustrates the maximum Krylov depth $\mathcal{M}$ (the total number of blocks) as a function of the system size $L$. We find that the Krylov depth of the Heisenberg model is always larger than that of the Ising model. This can be understood from the perspective of symmetry. The isotropic Heisenberg interaction commutes with the generator $\sum_{j}\hat{X}_{j}$, thereby preserving the associated $U(1)$ charge (i.e., the total $X$-magnetization). Consequently, repeated commutations with $\hat{H}_{\text{Heisenberg}}$ predominantly drive the growth of the operator size (Pauli weight) while remaining strictly confined within a specific charge sector. Only the transverse field $\hat{H}_{Z}$ fails to commute with this generator and consequently induces transitions between different charge sectors. This orthogonal interplay forces the operator growth to explore an effectively two-dimensional grid, spanned by the operator size and the $U(1)$ charge, which necessitates a greater depth $\mathcal{M}$ to fully explore the operator space.

    Having established the global spanning properties, we now shift our focus to the efficiency of generating specific local operators. Unless otherwise specified, all subsequent numerical results assume a system size of $L=6$. Fig.~\ref{fig:2}(b) evaluates the minimum block index $J$ required to access single-qubit Pauli operators $\hat{X}_i, \hat{Y}_i,$ and $\hat{Z}_i$. This metric essentially quantifies the inherent algebraic difficulty of extracting local control from purely global Hamiltonian dynamics. To further decode the internal structure of the block Krylov basis, Figs.~\ref{fig:2}(c) and \ref{fig:2}(d) map the minimum operator size against both the Krylov depth and the intra-layer index. Explicitly, by expanding any operator $\hat{A}$ in the Pauli basis as $\hat{A}=\sum_{P}c_{P}\hat{P}$, we denote the size $S(\hat{P})$ as the number of non-identity Pauli matrices within the string $\hat{P}$. The minimum size of $\hat{A}$, defined as $S_{\min}(\hat{A})=\min\{S(\hat{P}) \mid c_{P}\neq0\}$, thus serves as a rigorous quantitative metric for operator growth. This provides a refined characterization of operator spreading in the block Krylov basis, resolving the difficulty of realizing operators of a given size.
    
    \emph{ \color{blue} Optimal control.--} We now investigate the relation between the Krylov complexity $\mathcal{K}$ and the physical resource cost for realizing the target Hamiltonian. To determine the cost, we employ gradient-based optimization to determine the minimal evolution time required to realize a high-fidelity approximation of the target unitary $\hat{\mathcal{U}}_{\text{target}}=e^{-i\hat{H}_{\text{target}}\tau}$. As in practical experimental realizations of the general evolution operator \eqref{eq:Ugeneral}, we divide the total evolution into small time steps with fixed duration $\delta t$, during which the control amplitudes $\bm{u}(t)$ are held constant. After $n$ steps, the total evolution operator is $\hat{\mathcal{U}}_{n}=\prod_{k=1}^{n}\exp[-i\hat{H}(\bm{u}_{k})\delta t]$. The total evolution time $T = n\delta t$ is then tuned by varying the number of steps $n$, which quantifies the complexity of realizing the target evolution. We employ the Gradient Ascent Pulse Engineering (GRAPE) algorithm \cite{KHANEJA2005296} to minimize the loss function $\text{Loss}=1-\mathcal{F}(\hat{\mathcal{U}}_{n},\hat{\mathcal{U}}_{\text{target}})$, calculated from the unitary fidelity
\begin{equation}
\begin{aligned}
	\mathcal{F}(\hat{U},\hat{V})=\mathcal{D}^{-2}\left|\text{tr}(\hat{U}^{\dagger}\hat{V}) \right|^{2}.
\end{aligned}	
\end{equation}
	 To enhance optimization efficiency, a warm-start strategy is adopted: the total duration $T$ is progressively increased while the iteration budget is kept fixed. The optimized pulse sequence obtained from a preceding shorter duration then serves as an adaptive initial guess for the subsequent longer sequence.

\begin{figure}[t]
        \centering
        \includegraphics[width=1\linewidth]{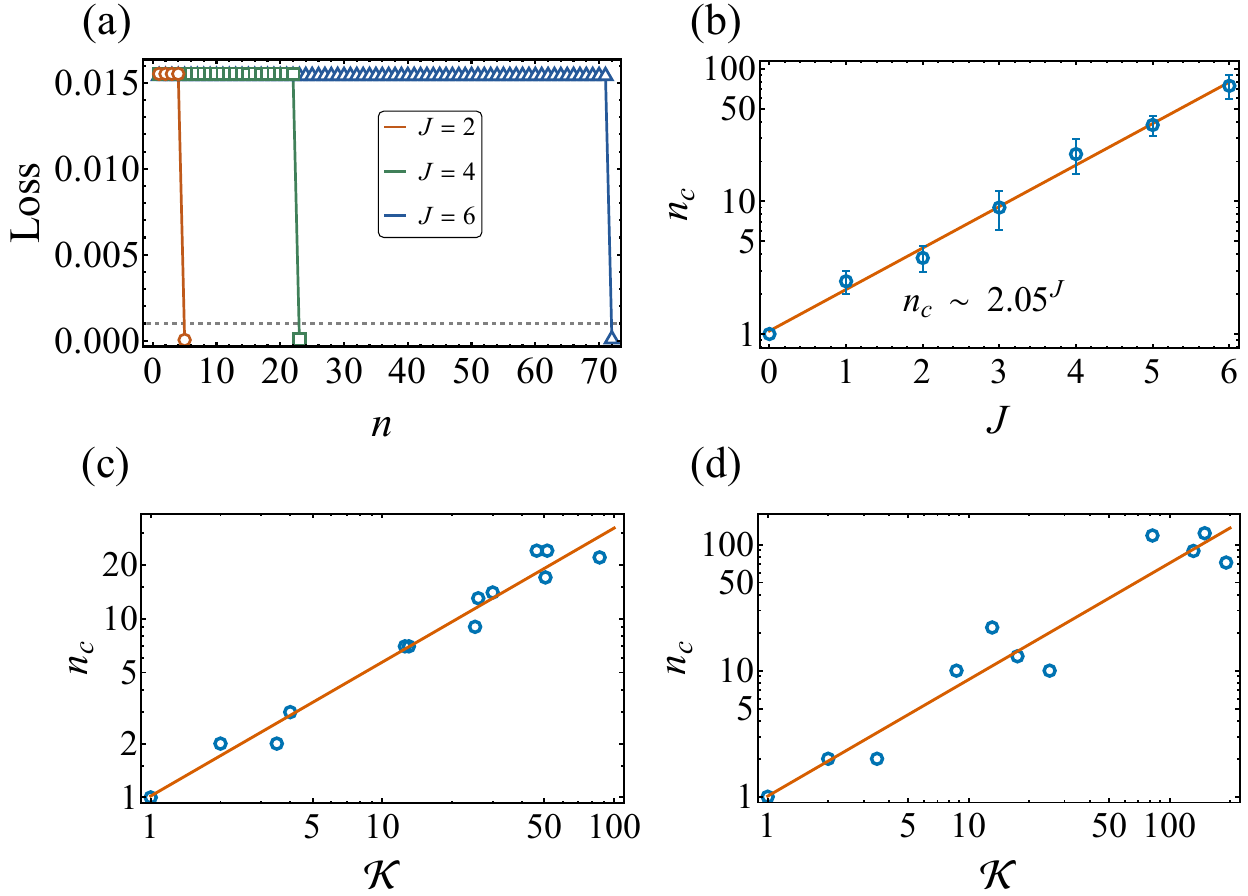}
        \caption{Synthesis time and Krylov complexity. (a) Optimization loss as a function of control steps $n$ for the Ising model. Target operators are randomly sampled from isolated Krylov layers $\Omega_J$. (b) Critical synthesis step $n_c$ required to achieve a target loss ($\epsilon=10^{-3}$) versus Krylov depth $J$. Error bars represent statistical averages over multiple random samples. (c)-(d) Scatter plots of $n_c$ versus the Krylov complexity $\mathcal{K}$ for Pauli string operators $\hat{P}$ with $S(\hat{P})\leq 2$ in the (c) Ising and (d) Heisenberg models.}
        \label{fig:3}
    \end{figure}
    
	To explicitly demonstrate the connection between this algebraic structure and the physical resource cost, we first investigate the synthesis of target operators residing in specific Krylov layers. Taking the Ising model as a prototypical example, Fig.~\ref{fig:3}(a) illustrates the optimization loss as a function of the number of control steps $n=T/\delta t$. In this analysis, we randomly sample target operators from the $J$-th Krylov block $\Omega_J$ and execute the gradient-based optimization. The resulting synthesis trajectories exhibit a clear hierarchy: operators situated at deeper Krylov layers (larger $J$) inherently necessitate a larger number of steps (larger $n$) to achieve a high-fidelity approximation. 

    Building upon these synthesis trajectories, we extract a quantitative scaling law for the physical resource cost. By setting the threshold with an acceptable loss of $\epsilon = 10^{-3}$, we identify the critical minimum step $n_c$ required to reach this threshold for each target operator. Fig.~\ref{fig:3}(b) presents this critical step $n_{c}$ plotted as a function of the Krylov depth $J$. The statistically averaged data reveals a clear exponential scaling of the required physical time with respect to the Krylov depth ($n_c \sim e^{\gamma J}$). This exponential resource growth is consistent with the asymptotic scaling of the theoretical synthesis complexity, $C_J \sim 2^J$.

    Having established the exponential scaling within individual Krylov blocks, we now turn our focus to target Hamiltonians generated by specific, single operators with small Pauli weights. These targets are chosen from the collection of low-weight Pauli strings, $\mathcal{S}=\{\hat{X}_{i},\hat{Y}_{i},\hat{Z}_{i}\}\cup\{\hat{Z}_{i}\hat{Z}_{j}\}_{i<j}$. Figs.~\ref{fig:3}(c) and \ref{fig:3}(d) present scatter plots of $n_c$ against the generalized Krylov complexity $\mathcal{K}$ for each of these distinct target Hamiltonians \footnote{The initial guess for the optimization is chosen randomly. Target operators that result in an exceedingly slow decrease of the loss function during single-shot optimization, likely due to being trapped in local minima, are excluded from the data presented in Figs.~\ref{fig:3}(c) and \ref{fig:3}(d).}. Notably, both the Ising and Heisenberg models exhibit a pronounced positive correlation between $n_c$ and $\mathcal{K}$, implying that $\mathcal{K}$ serves as a robust and universal proxy for the operational costs.

    \begin{figure}[t]
        \centering
        \includegraphics[width=1\linewidth]{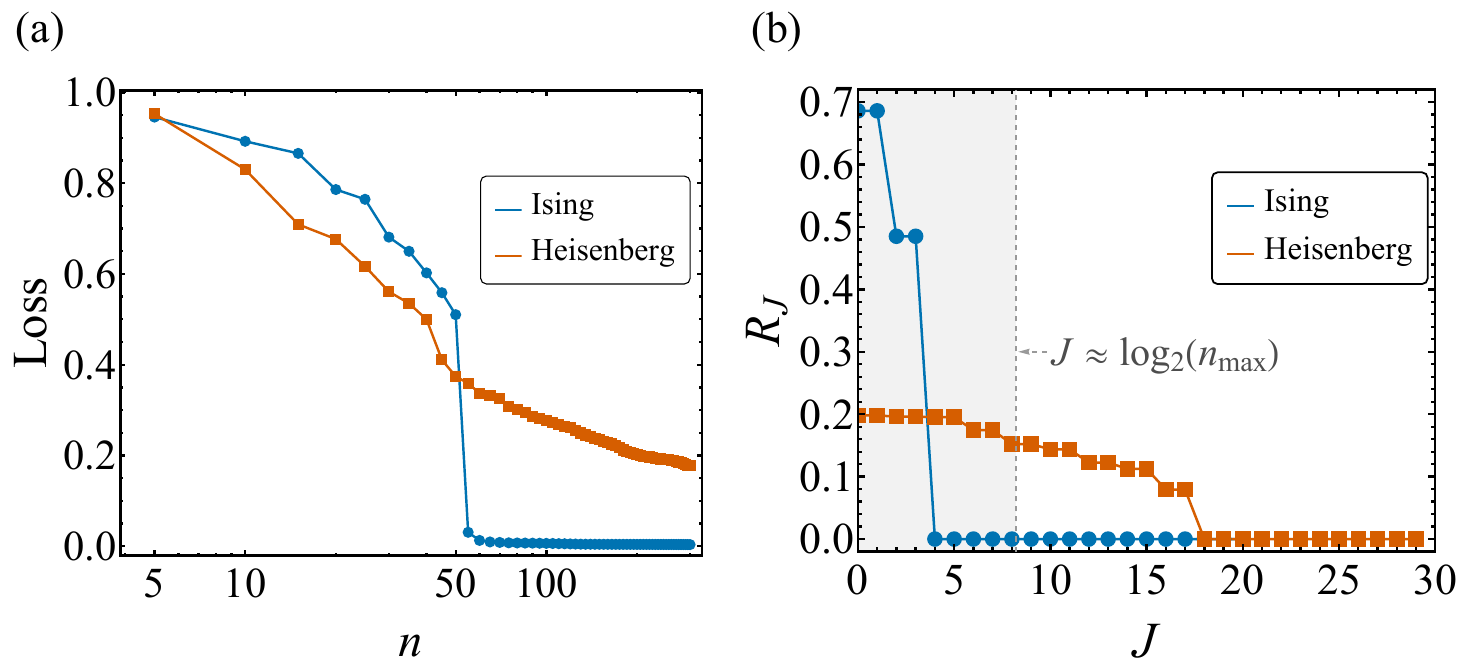}
        \caption{Synthesis trajectories and residual Krylov weights. (a) Optimization loss versus control steps $n$ for synthesizing the XXZ Hamiltonian using Ising and Heisenberg models. (b) The residual weight $R_m$, defined as the portion of the target Hamiltonian residing beyond Krylov depth $J$. The gray area corresponds to the region in (a). The dashed line is determined by $J \approx \log_2(n_{\max})$ with $n_{\max} = 300$.}
        \label{fig:4}
    \end{figure}
    
     Finally, we show that the block Krylov structure also governs the accuracy of the evolution for a given evolution time $T$. We choose a representative model of practical relevance in quantum many-body physics, the XXZ spin chain. The target Hamiltonian is given by
    \begin{equation}
\begin{aligned}
    \hat{H}_{\text{XXZ}}&=\sum_{i=1}^{L-1}\left( \hat{X}_{i}\hat{X}_{i+1}+\hat{Y}_{i}\hat{Y}_{i+1}+\Delta\hat{Z}_{i}\hat{Z}_{i+1} \right),
\end{aligned}
\end{equation}
    where the anisotropy parameter is set to $\Delta=1.5$. Fig.~\ref{fig:4}(a) shows the minimal achievable loss as a function of the total steps $n=T/\delta t$ when implemented by Ising and Heisenberg models, respectively, with an optimization budget of $N=20000$. To further elucidate the role of the algebraic structure, we calculate the residual weight beyond a given depth $J$
    \begin{equation}
        R_{J}=\sum_{m>J,n_{m}}\frac{\left| \langle\hat{\Omega}_{m,n_{m}}|\hat{H}_{\text{target}}\rangle \right|^{2}}{\langle\hat{H}_{\text{target}}|\hat{H}_{\text{target}}\rangle},
    \end{equation}
    which quantifies the fraction of the target Hamiltonian residing in the deeper sectors of the block Krylov basis. The behavior of $R_{J}$ for the XXZ model is shown in Fig.~\ref{fig:4}(b). The strong correlation between the synthesis trajectories and the residual weights $R_{J}$ implies that the simulation accuracy achieved for fixed $n$ is fundamentally governed by the structural representation of the target Hamiltonian in the block Krylov basis, where the maximal evolution time is directly related to the maximal Krylov depth.

    \emph{ \color{blue} Discussions.--} In this Letter, we establish a quantitative framework connecting Krylov complexity to the practical resource costs of quantum simulation. By generalizing the Krylov basis to a block structure, we validate the relevance of this measure by comparing it with gradient-based optimizations across various spin models. We observe that target Hamiltonians residing deep within the Krylov basis consistently require longer evolution times to synthesize with high fidelity. This firmly establishes the depth and distribution of the target within the block Krylov subspace as a strong predictor for the physical time cost in analog quantum simulators, offering a concrete tool for hardware-efficient pulse design.

   We conclude with several remarks on future directions. First, we have not yet explored the scaling behavior of the Lanczos coefficients \cite{Parker:2018yvk}, which are central to conventional studies of Krylov complexity. Understanding the structure and scaling of their block counterparts will be essential for a complete characterization of the simulator’s underlying algebraic structure. Second, a concrete measure of practical cost enables the reverse engineering of experimental platforms that can efficiently realize the target dynamics, potentially with the aid of artificial intelligence. Finally, it is important to understand how noise-induced errors affect the fidelity of the evolution, and in particular how these effects can be characterized by observables in Krylov space. This is both a fundamental and practical question for assessing the robustness of the simulation protocol.

    \vspace{5pt}
    \textit{Acknowledgement.}
    We thank Langxuan Chen and Hong-Ye Hu for helpful discussions. This project is supported by the Shanghai Rising-Star Program under grant number 24QA2700300, the NSFC under grant 12374477, and the Quantum Science and Technology-National Science and Technology Major Project 2024ZD0300101, and the Xuemin Fellow (Xuemin Institute of Advanced Studies, Fudan University).

\bibliography{main.bbl}
	
\end{document}